\documentclass[aps,prl,reprint]{revtex4-1}

\usepackage{lmodern}
\usepackage{braket}
\usepackage[usenames,dvipsnames]{pstricks}
\usepackage{epsfig}
\usepackage{pst-grad}
\usepackage{hyperref}

\begin{document}

\title{Quantum cost for sending entanglement}

\author{Alexander Streltsov}

\email{streltsov@thphy.uni-duesseldorf.de}

\author{Hermann Kampermann}

\author{Dagmar Bru\ss}

\affiliation{Institut f\"ur Theoretische Physik III, Heinrich-Heine-Universit\"at D\"usseldorf, D-40225 D\"usseldorf, Germany}

\maketitle

{\bf Establishing quantum entanglement between two distant parties is an essential
step of many protocols in quantum information processing 
\cite{Nielsen2000, Bruss2007}.
One possibility for providing long-distance  entanglement is to create
an entangled composite state within a lab and then physically send one subsystem to a distant
lab. However, is this the ``cheapest'' way? Here, we investigate the 
minimal ``cost''  that is necessary for establishing a certain amount of entanglement
between two distant parties.
We prove that this cost is intrinsically quantum,
and is specified by quantum correlations. 
Our results provide an optimal protocol for 
entanglement distribution and  show that quantum correlations are the
essential resource for this task. 
}


Imagine that one wants to send a letter in the old-fashioned way. 
The postage cost that the sender has to 
invest depends on the amount of the transmitted substance,
quantified by the weight of the letter. If the receiver had
already provided some pre-paid envelope, the sender may have to 
add an appropriate stamp if he/she wants to send a heavier letter.
Naturally, the allowed weight of the letter is smaller or equal to
a limit which is linked to the total postage.

Now, imagine that a sender wants to send quantum entanglement 
to a receiver. How does the cost that the sender has to
invest depend on the amount of entanglement sent, quantified by
some entanglement measure? Is this cost reduced when sender and
receiver already shared some pre-established entanglement?
And what is the nature of this cost - can one pay in classical
quantities, or does one have to invest a quantum cost?

One might be tempted to consider these questions and their
answers as obvious matters. However, quantum mechanics has
often surprised us with puzzling features: 
Counterintuitively, as shown in 
\cite{Cubitt2003}, {\em separable} states (i.e. states without
entanglement) can be used to distribute entanglement.  
What is then the resource that makes this process possible
and enables entanglement distribution without actually sending
an entangled state?

\begin{figure}
\scalebox{1} 
{
\definecolor{color1}{rgb}{0.35294117647058826,0.592156862745098,1.0}
\definecolor{color3}{rgb}{0.058823529411764705,0.9333333333333333,0.42745098039215684}
\definecolor{color2}{rgb}{0.9568627450980393,0.8431372549019608,0.054901960784313725}
\definecolor{color4}{rgb}{0.9568627450980393,0.8431372549019608,0.054901960784313725}

\begin{pspicture}(0,-5.7)(5.5,0.5)
\psdots[dotsize=0.5,linecolor=color1](0,0)
\rput(0,0){$A$}
\psdots[dotsize=0.5,linecolor=color3](5,0)
\rput(5,0){$B$}
\psdots[dotsize=0.5,linecolor=color1](0,-0.7)
\rput(0,-0.7){$C$}
\psline[linewidth=0.25cm,linecolor=color2](0.2,-0.7)(4,-0.7)
\psline[linewidth=0.25cm,linecolor=color4](4,-0.7)(4.8,-0.7)
\psarc[linestyle=none,linecolor=color2,fillstyle=solid,fillcolor=color2](0,-0.7){0.25}{-60}{60}
\psarc[linestyle=none,linecolor=color4,fillstyle=solid,fillcolor=color4](5,-0.7){0.25}{120}{-120}
\rput(2.5,-0.2){Initial setup}

\psdots[dotsize=0.5,linecolor=color1](0,-2)
\rput(0,-2){$A$}
\psdots[dotsize=0.5,linecolor=color3](5,-2)
\rput(5,-2){$B$}
\psline[linewidth=0.25cm,linecolor=color2](0.2,-2.7)(4,-2.7)
\psline[linewidth=0.25cm,linecolor=color4](4,-2.7)(4.8,-2.7)
\psarc[linestyle=none,linecolor=color2,fillstyle=solid,fillcolor=color2](0,-2.7){0.25}{-60}{60}
\psarc[linestyle=none,linecolor=color4,fillstyle=solid,fillcolor=color4](5,-2.7){0.25}{120}{-120}
\psdots[dotsize=0.5,linecolor=color2](2.5,-2.7)
\rput(2.5,-2.7){$C$}
\rput(2.5,-2.2){Transmission process}

\psdots[dotsize=0.5,linecolor=color1](0,-4)
\rput(0,-4){$A$}
\psdots[dotsize=0.5,linecolor=color3](5,-4)
\rput(5,-4){$B$}
\psdots[dotsize=0.5,linecolor=color3](5,-4.7)
\rput(5,-4.7){$C$}
\psline[linewidth=0.25cm,linecolor=color2](0.2,-4.7)(4,-4.7)
\psline[linewidth=0.25cm,linecolor=color4](4,-4.7)(4.8,-4.7)
\psarc[linestyle=none,linecolor=color2,fillstyle=solid,fillcolor=color2](0,-4.7){0.25}{-60}{60}
\psarc[linestyle=none,linecolor=color4,fillstyle=solid,fillcolor=color4](5,-4.7){0.25}{120}{-120}
\rput(2.5,-4.2){Final setup}

\psline[linewidth=0.04cm,linestyle=dotted, linecolor=color1](1,0.25)(1,-5.7)
\rput(0,-5.5){Alice's lab}
\psline[linewidth=0.04cm,linestyle=dotted, linecolor=color3](4,0.25)(4,-5.7)
\rput(5,-5.5){Bob's lab}

\end{pspicture} 
}
\caption{\label{fig1} Entanglement distribution between Alice and Bob. Blue
circles illustrate particles which belong to Alice, green circles
belong to Bob. The upper figure shows the initial setup
before the transmission: Alice holds the particles $A$ and $C$,
while Bob is in possession of the particle $B$. The middle figure shows the transmission process: Alice uses
a quantum channel (yellow) to send $C$ to Bob. The final situation
is shown in the lower figure. See also main text.
}
\end{figure}
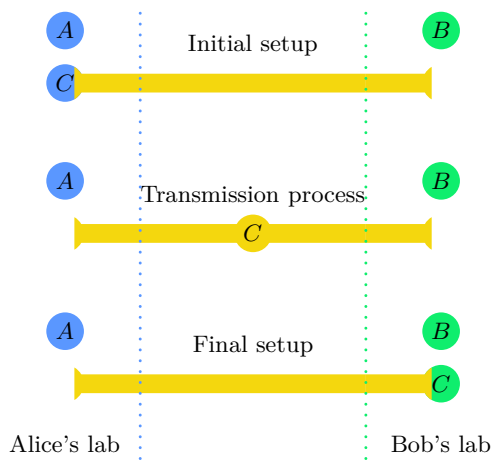

In order to address this question in a well-defined and quantitative way
we will consider the following
setting, see Fig. \ref{fig1}: the sender is called Alice ($A$), and the distant receiver
Bob ($B$). Each of them has a quantum particle in his/her possession.
In addition, they have a third quantum particle or
ancilla ($C$) available, which is at the beginning
located in Alice's lab, and then sent (via a noiseless quantum channel) to
Bob's lab. This is a general model for any interaction:
One can consider the particle $C$ as the intermediate
particle that realises the global interaction between $A$ and $B$. A similar scenario was also considered in a different context in \cite{Devetak2008,Yard2009}.

Initially, the total joint quantum state  may or may not carry entanglement. In the
following we will be only interested in bipartite entanglement,
i.e. two out of the three particles $A$, $B$ and $C$ are
grouped together. We quantify the initial entanglement between 
$AC$ and $B$ as $E^{AC|B}$,
and the final entanglement, after sending $C$ to Bob,  as 
$E^{A|BC}$. As a quantifier of entanglement we will first use the \emph{relative entropy
of entanglement}, which is a well established and widely studied measure
of entanglement for mixed states \cite{Vedral1997,Horodecki2009}. It is defined as the minimal relative entropy $S\left(\rho||\sigma\right)=\mathrm{Tr}\left[\rho\log\rho\right]-\mathrm{Tr}\left[\rho\log\sigma\right]$
between the given state $\rho^{XY}$ for two parties $X$ and $Y$
and the set of separable states ${\cal S}$:
\begin{equation}
E^{X|Y}\left(\rho^{XY}\right)=\min_{\sigma^{XY}\in{\cal S}}S\left(\rho^{XY}||\sigma^{XY}\right).\label{eq:Entanglement}
\end{equation}
Besides the fact that the relative entropy plays a crucial role in quantum information theory \cite{Vedral2002}, the significance of the relative entropy of entanglement is also provided by its close relation to the distillable entanglement \cite{Horodecki2000}.

In a naive approach to our original question, namely determining in a quantitative way
the cost for sending a certain amount of 
entanglement,  a natural conjecture would be the inequality 
$Q^{C|AB} \geq E^{A|BC} - E^{AC|B}$, where $Q$ denotes a yet undefined kind of correlations. This inequality can be interpreted as follows: If  initially Alice and Bob share some pre-established entanglement, quantified by $E^{AC|B}$,
and wish to achieve  final entanglement of $E^{A|BC}$ between them, the
ancilla $C$, sent from Alice to Bob,  needs to carry at least an amount of correlations given by the difference
of final and initial entanglement. This inequality quantifies the intuition, 
that entanglement distribution does not come for free, but always requires to invest some correlations.
In other words, $Q^{C|AB}$ could be interpreted as the "cost" for sending the entanglement $E^{A|BC} - E^{AC|B}$.
Quite surprisingly, it is \emph{not} the entanglement between $C$ and $AB$, which plays a crucial role here: as was demonstrated in \cite{Cubitt2003}, all steps of the protocol can be successfully implemented without any entanglement between $C$  and the rest of the system. In other words, if some inequality of the conjectured form
exists, the quantity $Q$ cannot be a measure of entanglement. However, does the fact that entanglement distribution
is possible via separable states mean that the ``cost'' for this protocol is of classical nature?
As we will show in the following, this is not the case: the cost for sending entanglement is of
quantum nature.

\begin{figure}
\begin{centering}
\includegraphics[width=0.69\columnwidth]{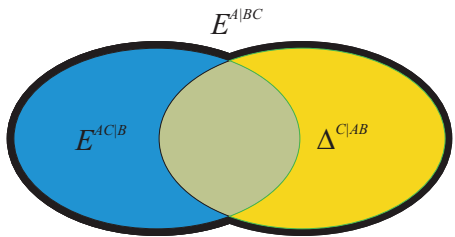}
\par\end{centering}

\caption{\label{fig:Venn} Illustration of the main result: The size of the blue area represents the entanglement between $AC$ and $B$, while the size of the yellow area represents the quantum correlations between $C$ and $AB$. The total area, enclosed by the black curve, represents the entanglement between $A$ and $BC$. One can  read off the main result: $E^{A|BC} \leq E^{AC|B} + \Delta^{C|AB}$.
}

\end{figure}

Even separable states, which by definition can be prepared locally with the help of classical
communication,  can carry quantum properties, i.e. they can be {\em quantum correlated}.
A composite quantum state is called strictly classically correlated if its correlations can
be described by a joint probability distribution for classical 
variables of the subsystems \cite{Piani2008}.
If this is not the case, quantum correlations are manifest in the state.
Recently, there has been much 
interest in characterising quantum correlations 
\cite{Zurek2000, Ollivier2001, Henderson2001, Oppenheim2002, Modi2010, Ferraro2010, Dakic2010}, in interpreting their occurrence in quantum
information protocols \cite{Zurek2003a, Madhok2011, Cavalcanti2011, Streltsov2011, Piani2011},
and in particular in determining their  role in quantum 
algorithms \cite{algorithms}, see also the feature article \cite{Merali2011} and the comprehensive review \cite{Modi2011}.
In the following we will quantify the amount of quantum correlations according to the thermodynamical approach presented in \cite{Oppenheim2002, Horodecki2005}. There the authors provided the notion of the information deficit: it quantifies the amount of information which cannot be localised by classical communication between two parties. If only one-way classical communication from party $X$ to party $Y$  is allowed, this leads to the \emph{one-way information deficit}: 
\begin{equation}
\Delta^{X|Y}\left(\rho^{XY}\right)=\min_{\left\{ \Pi_{i}^{X}\right\} }S\left(\rho^{XY}||\sum_{i}\Pi_{i}^{X}\rho^{XY}\Pi_{i}^{X}\right),\label{eq:Discord}
\end{equation}
where the minimisation is done over all local von Neumann measurements
$\left\{ \Pi_{i}^{X}\right\}$ on subsystem $X$.

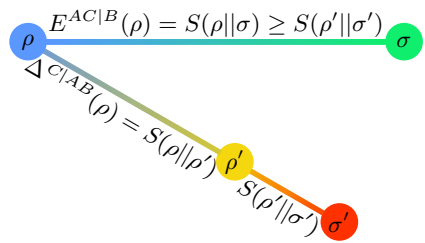
\begin{figure}
\scalebox{1} 
{
\definecolor{color1}{rgb}{0.35294117647058826,0.592156862745098,1.0}
\definecolor{color2}{rgb}{0.058823529411764705,0.9333333333333333,0.42745098039215684}
\definecolor{color3}{rgb}{0.9568627450980393,0.8431372549019608,0.054901960784313725}
\definecolor{color4}{rgb}{1.0,0.19215686274509805,0.0}

\begin{pspicture}(0,-2.5)(5.5,0.5)
\psdots[dotsize=0.5,linecolor=color1](0,0)
\rput(0,0){$\rho$}
\psdots[dotsize=0.5,linecolor=color2](5,0)
\rput(5,0){$\sigma$}
\psframe[linestyle=none,dimen=outer,fillstyle=gradient,gradlines=2000,gradbegin=color1,gradend=color2,gradmidpoint=1.0,gradangle=90.0](0.23,0.05)(4.77,-0.05)

\psdots[dotsize=0.5,linecolor=color3](2.75,-1.6)
\rput(2.75,-1.6){$\rho'$}
\rput{-30}{\psframe[linestyle=none,dimen=outer,fillstyle=gradient,gradlines=2000,gradbegin=color1,gradend=color3,gradmidpoint=1.0,gradangle=90.0](0.23,0.05)(2.97,-0.05)}

\psdots[dotsize=0.5,linecolor=color4](4.14,-2.4)
\rput(4.14,-2.4){$\sigma'$}
\rput{-30}(2.75,-1.6){\psframe[linestyle=none,dimen=outer,fillstyle=gradient,gradlines=2000,gradbegin=color3,gradend=color4,gradmidpoint=1.0,gradangle=90.0](0.23,0.05)(1.375,-0.05)}

\rput(2.5,0.25){$E^{AC|B}(\rho) = S(\rho||\sigma) \geq S(\rho'||\sigma')$}
\rput{-30}(1.25,-1){$\Delta^{C|AB}(\rho) = S(\rho||\rho')$}
\rput{-30}(3.33,-2.21){$S(\rho'||\sigma')$}

\end{pspicture} 
}

\caption{\label{fig:proof}Proof of the main result in Eq. (\ref{inequality}): The separable state $\sigma$
(green circle) is the closest separable state to the given state $\rho$
(blue circle). $ $The measured state $\rho'=\sum_{i}\Pi_{i}^{C}\rho\Pi_{i}^{C}$
(yellow circle) is defined such that $\Delta^{C|AB}\left(\rho\right)=S\left(\rho||\rho'\right)$. Application
of the same measurement on $\sigma$ gives the state $\sigma'=\sum_{i}\Pi_{i}^{C}\sigma\Pi_{i}^{C}$
(red circle). The states $\rho$, $\rho'$ and $\sigma'$ lie on a
straight line, for details see main text.}
\end{figure}

We will show in the following that the measure defined in Eq. (\ref{eq:Discord}) quantifies 
the cost discussed above, thus revealing the fundamental role of quantum correlations as 
a resource for the distribution of entanglement:
\begin{equation}
 \Delta^{C|AB} \geq E^{A|BC} - E^{AC|B}, 
\label{inequality}
\end{equation}
where the entanglement measure $E^{X|Y}$ was defined in Eq. (\ref{eq:Entanglement}).
This inequality is our central result; we will discuss its meaning and implications 
below. We point out that this inequality holds for any dimension of the three subsystems, see Fig. \ref{fig:Venn} for illustration. The main idea of the proof of Eq. (\ref{inequality}) is sketched in Fig. \ref{fig:proof}. We
name the state $\sigma$ to be the closest separable state to
$\rho$, i.e.  $E^{AC|B}\left(\rho\right)=S\left(\rho||\sigma\right)$.
We then
consider the local measurement $\{\Pi_{i}^{C}\}$ on particle $C$ that minimises the relative entropy of the
resulting state $\rho'$ with respect to the original $\rho$, i.e. 
$\rho'=\sum_{i}\Pi_{i}^{C}\rho\Pi_{i}^{C}$
such that $\Delta^{C|AB}\left(\rho\right)=S\left(\rho||\rho'\right)$.
In Fig. \ref{fig:proof} we also show the state $\sigma'=\sum_{i}\Pi_{i}^{C}\sigma\Pi_{i}^{C}$,
which results from the application of the same measurement on the
state $\sigma$. It is crucial to note that the three states $\rho$,
$\rho'$ and $\sigma'$ lie on a straight line, as shown in Fig. \ref{fig:proof}:
\begin{equation}
S\left(\rho||\sigma'\right)=S\left(\rho||\rho'\right)+S\left(\rho'||\sigma'\right).\label{eq:line}
\end{equation}
For proving this equality it is enough to show the relations $\mathrm{Tr}\left[\rho\log\rho'\right]=\mathrm{Tr}\left[\rho'\log\rho'\right]$
and $\mathrm{Tr}\left[\rho\log\sigma'\right]=\mathrm{Tr}\left[\rho'\log\sigma'\right]$,
then Eq. (\ref{eq:line}) immediately follows. 
These two equalities can be shown in a straight-forward way, by using the idempotent property of the projectors,
the cyclic invariance of the trace, and the fact that the projectors $\Pi_{i}^{C}$ sum
up to the identity.

The final ingredient in the proof of Eq. (\ref{inequality}) is the fact that the relative entropy
does not increase under quantum operations \cite{Lindblad1975,Wehrl1978,Vedral1997}: $S\left(\Lambda\left(\rho\right)||\Lambda\left(\sigma\right)\right)\leq S\left(\rho||\sigma\right)$,
and thus $S\left(\rho'||\sigma'\right)\leq S\left(\rho||\sigma\right)$.
Inserting this into Eq. (\ref{eq:line})
implies the inequality $S\left(\rho||\sigma'\right)\leq\Delta^{C|AB}\left(\rho\right)+E^{AC|B}\left(\rho\right).$
To complete the proof of Eq. (\ref{inequality}), we notice that
the state $\sigma'$ is a tripartite fully separable state, and thus
gives an upper bound on the entanglement $E^{A|BC}\left(\rho\right)\leq S\left(\rho||\sigma'\right)$.

The techniques presented above can also be applied to a more general
measure of entanglement, where the relative entropy $S(\rho_1||\rho_2)$ is replaced
- both for the entanglement measure and the quantum correlation measure - 
by a general distance $D(\rho_1,\rho_2)$. We only demand that $D$ has the following
two properties: 
\begin{itemize}
\item $D$ does not increase under all quantum operations,
\item $D$ satisfies the triangle inequality.
\end{itemize}
Then Eq. (\ref{eq:line}) becomes an inequality: $D\left(\rho,\sigma'\right)\leq D\left(\rho,\rho'\right)
+D\left(\rho',\sigma'\right)$,
and the proof of Eq. (\ref{inequality}) follows from the same
arguments as above. Well-known and frequently used examples for distances that fulfil these two properties 
\cite{Nielsen2000} are e.g.
the trace distance, defined as 
$D_t(\rho_1,\rho_2) = \frac{1}{2}\mathrm{tr}|\rho_1 -\rho_2|$ and the Bures distance \cite{Uhlmann1992}, defined as
$D_B(\rho_1,\rho_2) = 2(1-\sqrt{F(\rho_1,\rho_2)})$,
with 
$F(\rho_1,\rho_2) = (\mathrm{tr}\sqrt{\sqrt{\rho_1}\rho_2\sqrt{\rho_1}})^2$.

Let us point out that our  main result in inequality (\ref{inequality}) 
can be alternatively 
seen as a restricting link between the correlation
properties of the three possible
bipartite splits of a tripartite quantum state in any dimension: 
the entanglement across one of the bipartite splits cannot be larger than the sum
of the entanglement across one of the other splits plus the quantum
correlations across the remaining split.
Thus, the  inequality (\ref{inequality})
 may be interpreted as a type
of ``monogamy'' relation between three entangled parties. 
This inequality also
holds for all permutations of the parties.
 By permuting the systems $A$ and $B$ in Eq. (\ref{inequality}), we obtain the generally valid inequality
\begin{equation}
E^{AC|B} - \Delta^{C|AB} \leq E^{A|BC} \leq E^{AC|B} + \Delta^{C|AB}. 
\label{generalineq}
\end{equation}
This inequality tells us, that the entanglement between $A$ and $BC$ is not  independent from the entanglement between $AC$ and $B$. In particular, in the case of vanishing quantum correlations, i.e.  $\Delta^{C|AB} = 0$,
 we immediately see that these two quantities are equal: $E^{A|BC}=E^{AC|B}$. -
We also note that for those
situations, where $\Delta^{C|AB}= E^{C|AB}$ - this happens
e.g. for the relative entropy when the state under consideration is pure - one arrives, using
all permutations of inequality (\ref{inequality}), at the 
triangle inequality $|E^{B|AC} - E^{C|AB}| \leq E^{A|BC} \leq E^{B|AC} + E^{C|AB}$.
However, we stress again that this symmetric inequality  is a special case
of the general inequality (\ref{generalineq}), and is valid only for certain classes of states.

We are now in position to answer the question posed in the first paragraph
of this paper: \emph{What is the cheapest way for distributing entanglement?}
In order to answer this question in full generality, we consider the
most general distribution protocol, which may contain $n$ uses of the quantum channel together with local operations and classical
communication between Alice and Bob. The amount
of entanglement sent in this process of \emph{entanglement
growing} cannot be larger than the total cost in the protocol:
\begin{equation}
E_{\mathrm{final}}-E_{\mathrm{initial}}\leq\sum_{i=1}^{n}\Delta_{i},\label{eq:nrounds}
\end{equation}
where $E_{\mathrm{initial}}$ and $E_{\mathrm{final}}$ is the amount of entanglement between Alice and Bob before and after the protocol, and  $\Delta_{i}$ is the amount of quantum correlations between
the sent particle and the remaining system in the $i$-th application
of the quantum channel.

In order to prove Eq. (\ref{eq:nrounds}), we first consider a protocol where the quantum channel is used once from Alice to Bob and once in the other direction, i.e. $n=2$. Suppose
that Alice and Bob start with a state $\rho_{1}$, the initial
entanglement is $E_{\mathrm{initial}}=E^{AC|B}\left(\rho_{1}\right)$.
After sending the particle $C$ to Bob the entanglement between the
two parties is given by $E^{A|BC}\left(\rho_{1}\right)$, and the
cost for this process is given by $\Delta^{C|AB}\left(\rho_{1}\right)$.
Now Alice and Bob locally act on their subsystems, and may additionally
communicate classically with each other, thus arriving at the final
state $\rho_{2}$ with the entanglement $E^{A|BC}\left(\rho_{2}\right)$.
In the final step of this single-round protocol Bob sends the particle
$C$ back to Alice, and the final entanglement is $E_{\mathrm{final}}=E^{AC|B}\left(\rho_{2}\right)$.
The corresponding cost for this final step is given by $\Delta^{C|AB}\left(\rho_{2}\right)$.
We will now show that the amount of entanglement sent in the total
process cannot be larger than the total cost: 
\begin{equation}
E_{\mathrm{final}}-E_{\mathrm{initial}}\leq\Delta^{C|AB}\left(\rho_{1}\right)+\Delta^{C|AB}\left(\rho_{2}\right).\label{eq:tworounds}
\end{equation}
This inequality can be seen by applying inequality (\ref{inequality}) to the two states
$\rho_{1}$ and $\rho_{2}$ independently, and considering the sum
of the both inequalities: $E^{AC|B}\left(\rho_{2}\right)-E^{A|BC}\left(\rho_{2}\right)+E^{A|BC}\left(\rho_{1}\right)-E^{AC|B}\left(\rho_{1}\right)\leq\Delta^{C|AB}\left(\rho_{2}\right)+\Delta^{C|AB}\left(\rho_{1}\right)$.
Note that the entanglement $E^{A|BC}\left(\rho_{2}\right)$ is not
larger than $E^{A|BC}\left(\rho_{1}\right)$, since the state $\rho_{2}$
results from the state $\rho_{1}$ after application of local operations
and classical communication. This proves the desired inequality (\ref{eq:tworounds}).
  To prove the general expression in Eq. (\ref{eq:nrounds}), we now suppose that the quantum channel is used $n$ times, where $n$ can be even or odd. We can define the states $\rho_1,\ldots,\rho_n$ in an analogous
 way as above. Using the same argumentation we arrive at Eq. (\ref{eq:nrounds}).

The result in Eq. (\ref{eq:nrounds}) can now be used to find the
most ``economic'' way to distribute entanglement. If Alice and Bob are told
to send a fixed amount of entanglement $E=E_{\mathrm{final}}-E_{\mathrm{initial}}$,
they can achieve this in the most economic way by choosing a protocol such
that the inequality (\ref{eq:nrounds}) becomes an equality. 
One possibility to achieve this is the well-known ``trivial'' one:
Alice locally prepares a pure state $\ket{\psi}^{AC}$
with entanglement $E=E^{A|C}$, and sends the particle $C$ to Bob.
However, this is not the only possibility: the inequality (\ref{eq:nrounds})
can also be satisfied {\em without} sending entanglement, see the example below.
If one considers entanglement to be an expensive resource, one may thus be able to 
distribute entanglement in a ``cheaper'' way by sending quantum correlations without entanglement.

The results presented in this work provide new powerful tools to understand and
quantify  entanglement as well as  quantum correlations. 
In this paragraph we will demonstrate how Eq. (\ref{inequality})
can be used to evaluate the entanglement and discord in the specific state $\eta$, which was used in \cite{Cubitt2003} to show that entanglement distribution with separable states is possible:
\begin{equation}
\eta=\frac{1}{3}\ket{\Psi_{GHZ}}\bra{\Psi_{GHZ}}+\sum_{i,j,k=0}^{1}\beta_{ijk}\Pi_{ijk}
\end{equation}
with $\ket{\Psi_{GHZ}}=\frac{1}{\sqrt{2}}\left(\ket{000}+\ket{111}\right)$,
$\Pi_{ijk}=\ket{ijk}\bra{ijk}$, and all $\beta$'s are zero apart
from $\beta_{001}=\beta_{010}=\beta_{101}=\beta_{110}=\frac{1}{6}$.
It was shown in \cite{Cubitt2003} that the entanglement is zero between two different
cuts: $E^{AC|B}=E^{AB|C}=0$. As an application of Eq. (\ref{inequality}) we will now prove that the remaining two quantities are equal: $E^{A|BC}\left(\eta\right)=\Delta^{C|AB}\left(\eta\right)=\frac{1}{3}$. This can be seen by considering the relative entropy between $\eta$
and the state $\eta'=\sum_{i}\Pi_{i}^{C}\eta\Pi_{i}^{C}$
with orthogonal projectors $\Pi_{i}^{C}=\ket{i}\bra{i}^{C}$ in the computational basis. It can
be verified by inspection that $S\left(\eta||\eta'\right)=\frac{1}{3}$, and thus $\Delta^{C|AB}\left(\eta\right)$ is not larger than $\frac{1}{3}$. 
On the other hand, the entanglement $E^{A|BC}\left(\eta\right)$ is bounded from below by
$\frac{1}{3}$. This follows from the two facts that the state $\eta$ can be used to distil Bell states
with probability $\frac{1}{3}$ \cite{Cubitt2003}, and that the relative entropy of entanglement is not smaller than the distillable entanglement \cite{Horodecki2000}.
- In this example, 
quantum correlations provide the most economic and cheapest resource
for entanglement distribution.

In conclusion, we have identified quantum correlations as the key resource for entanglement distribution.
They quantify the quantum cost that one has to invest for increasing the entanglement 
between two distant parties. Explicitly, we proved that the entanglement between two
parties cannot grow more than the amount of quantum correlations which the
particle carries that mediates the interaction between the two parties. Our result is completely
general and is valid regardless of the particular realisation of the protocol. Thus it provides
a fundamental connection between quantum entanglement on one side and quantum 
correlations on the other side. 
Since the study of quantum correlations is believed to be important for understanding the power 
of quantum computers, our results may find applications far beyond the scope of this work.

We thank Otfried G\"uhne for helpful discussions. AS  thanks Kavan Modi, Hugo Cable and Vlatko Vedral for organising the Quantum Discord Workshop 2012 in Singapore, where he had enlightening discussions with Gerardo Adesso, Wojciech H. Zurek, and Animesh Datta. 
We acknowledge financial support from Deutsche Forschungsgemeinschaft (DFG) and ELES.

Note added: During the completion of this work we became aware of independent related work by T. K. Chuan et al. in \cite{Chuan2012}.

\bibliographystyle{apsrev4-1}
\bibliography{literature}

\end{document}